\documentclass[reprint,aps,prc,twocolumn,superscriptaddress,floatfix,10pt]{revtex4-2}
\usepackage{bm}
\usepackage{amssymb,amsbsy,amsmath,amsfonts}
\usepackage{graphicx}
\usepackage{color}
\usepackage{xcolor}
\usepackage{booktabs}
\usepackage[colorlinks,allcolors=blue]{hyperref}
\makeatletter
\def\NAT@def@citea{\def\@citea{\NAT@separator}}
\makeatother

\usepackage{orcidlink}

\begin{document}
\title{Radii of light nuclei from the Jacobi No-Core Shell Model}

\author{Xiang-Xiang Sun\orcidlink{0000-0003-2809-4638}}
\email{x.sun@fz-juelich.de}
\affiliation{Institute for Advanced Simulation (IAS-4), Forschungszentrum J\"{u}lich, D-52425 J\"{u}lich, Germany}

\author{Hoai Le\orcidlink{0000-0003-1776-9468}}
\email{h.le@fz-juelich.de}
\affiliation{Institute for Advanced Simulation (IAS-4), Forschungszentrum J\"{u}lich, D-52425 J\"{u}lich, Germany}

\author{Ulf-G. Mei{\ss}ner\orcidlink{0000-0003-1254-442X}}
\email{meissner@hiskp.uni-bonn.de}
\affiliation{Helmholtz-Institut~f\"{u}r~Strahlen-~und~Kernphysik~and~Bethe~Center~for~Theoretical~Physics, Universit\"{a}t~Bonn,~D-53115~Bonn,~Germany} 
\affiliation{Institute for Advanced Simulation (IAS-4), Forschungszentrum J\"{u}lich, D-52425 J\"{u}lich, Germany}
\affiliation{Peng Huanwu Collaborative Center for Research and Education, Beihang University, Beijing 100191, China}
\affiliation{CASA, Forschungszentrum J\"{u}lich, 52425 Ju\"{u}lich, Germany}

\author{Andreas Nogga\orcidlink{0000-0003-2156-748X}}
\thanks{Corresponding author: a.nogga@fz-juelich.de}
\affiliation{Institute for Advanced Simulation (IAS-4), Forschungszentrum J\"{u}lich, D-52425 J\"{u}lich, Germany}
\affiliation{CASA, Forschungszentrum J\"{u}lich, 52425 Ju\"{u}lich, Germany}

\date{\today}
\begin{abstract}
Accurately determining the size of the atomic nucleus with
realistic nuclear forces is a long outstanding issue of
nuclear physics. 
The no-core shell model (NCSM), one of the powerful 
\textit{ab initio} methods for nuclear structure, can achieve accurate energies of light nuclei. The extraction of converged radii is more difficult.
In this work, we present a novel method to effectively extract the radius of light nuclei by restoring the 
long-range behavior of densities from NCSM calculations.
The correct large distance asymptotic of two-body relative densities are deduced based on the NCSM densities in limited basis size. 
The resulting radii using the corrected densities show a nice convergence. The root-mean-square matter and charge radii of $^{4,6,8}$He and $^{6,7,8}$Li can be accurately obtained 
based on Jacobi-NCSM calculations with the high-precision chiral two-nucleon and three-nucleon forces combined with this new method.  
Our method can be straightforwardly extended to other \textit{ab initio} calculations, potentially providing a better description of nuclear sizes with realistic nuclear forces.

\end{abstract}
\maketitle
\section{Introduction}

The radius is one of the most important properties of the atomic nucleus.
At present, with the increase in high-performance
computing resources, advanced nuclear \textit{ab initio} approaches, including the
no-core shell model (NCSM), 
quantum Monte-Carlo,
in-medium similarity-renormalization-group (IMSRG),
coupled-cluster theory, and 
nuclear lattice simulations
\cite{Epelbaum:2008ga,Machleidt:2011zz,Barrett:2013nh,
Hebeler:2015hla,Hergert:2015awm,Lahde:2019npb,
Lee:2020meg,Hammer:2012id,Carlson:2014vla,Hagen:2013nca,Machleidt:2024bwl,Elhatisari:2022zrb,Maris:2020qne,LENPIC:2022cyu},
high-precision nuclear forces 
have been widely used to study both nuclear structure and reactions.
With the help of unitary transformations like $V_{\mathrm{low-}k}$
\cite{Bogner:2003wn}, 
the unitary correlation operator method (UCOM) 
\cite{Roth:2010bm}
or the similarity renormalization group (SRG)
\cite{Bogner:2006pc,Jurgenson:2009qs}, the interactions 
are systematically softened  and 
most \text{ab initio} calculations  can achieve
converged energies using a limited basis size.   
The results 
based on  the softened chiral two-nucleon (NN) and three-nucleon (3N) forces  are generally consistent with experiments \cite{LENPIC:2022cyu}. 
However, the predicted radii do not provide accurate descriptions for precision measurements of nuclear charge radii, 
which remains a challenge for 
nuclear structure theories \cite{Hergert:2016etg,Bissell:2016vgn,Lapoux:2016exf,GarciaRuiz:2016ohj,Koszorus:2020mgn,Kaur:2022yoh,Maris:2020qne,LENPIC:2022cyu,Maris:2023esu}, see, however, Ref.~\cite{Elhatisari:2022zrb}.

The nuclear many-body wave functions and the corresponding densities from \textit{ab initio} NCSM or IMSRG using the SRG-evolved realistic nuclear interaction have two significant limitations. 
First, while applying the softened interactions facilitates the convergence of energy calculations,  
the unitary transformation alters the 
wave function at the small distance (high momentum) regime.
While such changes in wave function are unobservable, 
it indicates that
the operators have to be consistently evolved to extract
the pertinent observables, 
which can be achieved by performing the SRG evolution of the operators
or directly getting the unitary transformation matrix \cite{Anderson:2010aq,Schuster:2014lga}. 
Its effects on nuclear size, a long-range observable, 
have been checked to be  small \cite{Schuster:2014lga, Hergert:2016etg,Miyagi:2019bkl} 
but it notably influences the short-range or
high-momentum nuclear density distribution
\cite{Neff:2015xda}.

Second, due to the limited 
model space size and the harmonic oscillator (HO) basis functions, the long-range part of the density falls as $e^{-\beta r^2}$ rather than the expected $e^{-\kappa r}$, where $\beta$ is related to the oscillator frequency and $\kappa$ is given by the binding momentum. 
For obtaining binding energies, convergence outside the range of the interaction is not required and, generally, convergence of the tail of the wave function is not achieved for this part of the wave function even when modern high-performance computing resources are used.  
As a result, the long-range observables, including roots-mean-square (rms) radius and $E2$ transitions and moments,  show a strong dependence on the cutoff of the HO basis size and its frequency (see investigations of light nuclei using NCSM or no-core configuration interaction calculations \cite{Bogner:2007rx,Cockrell:2012vd,Maris:2013poa,Heng:2016umo,Shin:2016poa,Choudhary:2020wgv,Caprio:2021umc,Rodkin:2022fus}), 
deviating from the usual monotonic convergence pattern observed in binding energy calculations with increasing basis.
Therefore, the ``crossover prescription'' 
\cite{Maris:2013poa, Caprio:2014iha},  
extrapolation procedures \cite{Furnstahl:2013vda,Furnstahl:2012qg,Forssen:2017wei, Rodkin:2022fus} 
and artificial neural networks (ANN) \cite{Negoita:2018kgi,Wolfgruber:2023ehw} 
have been applied to extract the radii in NCSM calculations.
It should be noted that this issue also persists even when employing bases with proper asymptotic behavior, such as the Coulomb-Sturmian basis \cite{Caprio:2014iha} or natural orbitals \cite{Constantinou:2016urz,Fasano:2021ahd,Tichai:2018qge}. 

Very recently, the \textit{ab initio} calculations using the higher-order chiral semilocal momentum-space (SMS) regularized NN forces (N$^4$LO$^+$) in combination with the 3N forces (3NFs) at N$^2$LO (SMS N$^4$LO$^+$+N$^2$LO) have proven to be able to describe the binding energies of light nuclei very well \cite{Maris:2020qne, LENPIC:2022cyu}. 
But the results indicate that radii of nuclei in the upper $p$-shell  
might be underpredicted in these calculations. It is still an open question whether the deviation from experiment can be resolved by a higher 
3NFs or taking NN electromagnetic current operators into account \cite{Filin:2020tcs} or whether an improved converges can 
resolve this problem at least partly. 
In this context, an application 
of an ANN for the extrapolation of the rms matter radius of light nuclei \cite{Wolfgruber:2023ehw} based on the same interactions is of 
high interest since it promises high accuracy results for this quantity. 
In this work,
focusing on the above-mentioned two issues with the nuclear radii computed within the NCSM framework,
we propose an alternative method to extract the nuclear radii based on the densities from the calculations using NCSM based on relative Jacobi coordinates (J-NCSM)
\cite{Liebig:2015kwa, Le:2020zdu, Le:2022ikc, Le:2021gxa, Le:2023bfj}
with 
SMS N$^4$LO$^+$+N$^2$LO interaction 
\cite{LENPIC:2022cyu}. 
This paper is organized as follows: 
in Sec.~\ref{Sec:II}, we briefly introduce the formalism. 
Then, 
in Sec.~\ref{Sec:III}, the results for the radii
of light nuclei are presented and discussed. 
Details on an SRG flow parameter dependence and the 
definition of improved densities are given in the Appendices~\ref{APP:A1} 
and \ref{APP:A2}. We summarize and put our work in perspective in Sec.~\ref{Sec:IV}.

\section{Theoretical framework}\label{Sec:II}
We start with a nuclear
many-body Hamiltonian containing the kinetic term, and chiral NN and 3N forces
\begin{equation}
H_0 = T + V^{2N} + V^{3N}.
\end{equation}
The strong repulsive core of realistic nuclear interaction
makes this Hamiltonian so hard to solve directly 
in nuclear many-body methods. 
In practical calculations, this interaction is softened 
via the unitary transformation, 
for example, the most commonly used
SRG transformation \cite{Bogner:2006pc,Jurgenson:2009qs},  
\begin{equation}
H_s^{} = U_s^{} H_0^{} U^{\dagger}_s 
= T_\mathrm{rel}^{} + V^{2N}_s +V^{3N}_s,
\end{equation}
where the SRG evolved forces $V_s$ and unitary transformation operators 
are governed by the flow equations 
\begin{equation}
\frac{dV_s}{ds} = [\eta_s,H_s], 
\end{equation}
and 
\begin{equation}
\frac{dU_s}{ds} = \eta_s U_s,
\label{eq:srg}
\end{equation}
where the most common generator of the transformation is $\eta_s=[T_\mathrm{rel},H_s]$ 
and 
$s$ is the flow parameter. In the following, we quantify the flow parameter with $\lambda = (s/m_N^2)^{-1/4}$ scaled by the nucleon mass $m_N$ since 
this quantity can be interpreted as an effective momentum 
cutoff \cite{Bogner:2006pc}.
The SRG flow equations with NN and 3N interactions are solved at the three-body level.
The SRG evolved interactions are the input of our J-NCSM calculations
from which we also get the nuclear many-body wave functions corresponding to
the SRG softened interaction, in the following  called the `low-resolution solution' and denoted as 
$|\Psi_\lambda\rangle$. In this solution, due to the unitary transformation,
the high-momentum information is encoded at low momenta different to 
the solution $|\Psi\rangle $ of the bare interactions.
In principle, all the observables should be calculated as 
\begin{equation}
\langle \hat{O} \rangle = \langle \Psi| \hat{O} | \Psi \rangle = 
 \langle \Psi_\lambda| \hat{O}_\lambda | \Psi_\lambda\rangle.
\end{equation}
In practice,
we know the wave function of SRG softened interaction from  and corresponding unitary 
transformation from, the observable can be obtained using 
\begin{equation}
\langle \hat{O} \rangle = \langle \Psi_\lambda^{}| U_\lambda^{}\hat{O} U_\lambda^\dagger  | \Psi_\lambda^{} \rangle.
\end{equation}

The rms matter radius $R_m$ and the rms point-proton radius $R_p$ or rms point-neutron radius $R_n$ can be calculated using \cite{Bacca:2012up}
\begin{eqnarray}
&& R^2_{m} = \frac{1}{A}\sum_{i}(\bm r_{i}-\bm R_\text{c.m.})^2, \\
&& R^2_{p} = \frac{1}{Z}\sum_{i}\frac{1+\tau_{z,i}}{2}(\bm r_{i}-\bm R_\text{c.m.})^2, \\
&& R^2_{n} = \frac{1}{N}\sum_{i}\frac{1-\tau_{z,i}}{2}(\bm r_{i}-\bm R_\text{c.m.})^2, 
\end{eqnarray}
where $\bm R_\text{c.m.}=\sum_i \bm r_{i}/A$, $\tau_{z,i}$ is the third component of the isospin of the nucleon $i$ (with $\tau_z=+1$ for protons and $\tau_z=-1$ for neutrons), and $A,N,Z$ label the number of nucleons, neutrons, and protons.

To apply the SRG transformation to the wave function or density, 
it is advantageous to reformulate the radius operator in terms of two-nucleon 
distance operators. This operator can  be achieved through the two-body matrix element defined in the normalization specified below as 
\cite{Caprio:2012rv, Caprio:2014iha}
\begin{eqnarray}
\label{eq:radius}
\left\langle R_m^2\right\rangle & 
=& \frac{Z(Z-1)}{2A^2}\langle R_\text{rel}^2 \rangle_{p p}+
\frac{N(N-1)}{2A^2}\left\langle R_\text{rel}^2\right\rangle_{n n} \nonumber \\
& & +
\frac{NZ}{A^2}\left\langle R_\text{rel}^2\right\rangle_{n p}, \nonumber \\ 
\left\langle R_p^2\right\rangle & 
=&\frac{(A+N)(Z-1)}{2A^2}\left\langle R_\text{rel}^2\right\rangle_{p p}+
\frac{N^2}{A^2}\left\langle R_\text{rel}^2\right\rangle_{n p} \nonumber \\
& & -\frac{N(N-1)}{2A^2}\left\langle R_\text{rel}^2\right\rangle_{n n} ,\nonumber \\ 
\left\langle R_n^2\right\rangle & 
=&-\frac{Z(Z-1)}{2A^2}\left\langle R_\text{rel}^2\right\rangle_{p p}+
\frac{Z^2}{A^2}\left\langle R_\text{rel}^2\right\rangle_{n p} \nonumber \\
& & +
\frac{(A+Z)(N-1)}{2A^2}\left\langle R_\text{rel}^2\right\rangle_{n n},
\end{eqnarray}
where the expectation value of these two-body operators can be calculated using the two-body relative densities from our J-NCSM wave functions. Our calculations are based on the two-nucleon transition densities $\rho_{\alpha' \alpha}^{M'M\,m_t}(p',p)$ introduced in \cite{Griesshammer:2020ufp} and available online at 
\cite{densdb:2025nuc}. The depend on the third component of isospin of the NN pair $m_t=-1,0,1$, the third component of total angular momentum of the  
in- and outgoing nucleus $M$ and $M'$, $M,M'=-J,\ldots,J$, where $J$ is the angular momentum of the considered state. 
The densities also depend on the magnitude of the momenta $p'$,$p$ and the NN partial waves of the pair $\alpha$ and $\alpha'$. Thereby, $\alpha$ refers to $(ls)j m_j \, t$, namely to the NN orbital relative orbital angular momentum $l$ coupling with the spin of the NN system $s$ to the total NN angular momentum $j$ and its third component $m_j$. In order to end up with densities for the relative distance a Fourier transformation needs to be performed. 
Already performing the sum over all NN partial waves, 
and averaging on the polarization of the nucleus, this results in the densities 
\begin{eqnarray}
\label{eq:dennorm}
    \rho_{m_t}(r) & =&  \frac{1}{2J+1} \sum_{M \alpha}
      \frac{2}{\pi}\int dp \, p^2 \int dp' \, p^{\prime 2} \ 
      j_l(pr)\, j_l(p'r) \nonumber \\ & & \hspace{2cm}\times\rho_{\alpha \alpha}^{MM\, m_t}(p',p)
      \end{eqnarray}
which are normalized to 
\begin{eqnarray}
    \sum_{m_t} \int dr \, r^2 \ \rho_{m_t}(r) = 1 \ .
\end{eqnarray}
The rms distance of $pp$, $nn$ and $np$ pairs is then obtained by 
\begin{equation}
\begin{split}
& \langle R_\text{rel}^2 \rangle_{p p} = \frac{A(A-1)}{Z(Z-1)}\int dr \, r^4 \rho_{m_t=1}(r) , \\
& \langle R_\text{rel}^2 \rangle_{n n} = \frac{A(A-1)}{N(N-1)}\int dr \, r^4 \rho_{m_t=-1}(r) ,\\
& \langle R_\text{rel}^2 \rangle_{n p} = \frac{A(A-1)}{2NZ} \int dr \, r^4 \rho_{m_t=0}(r) \ .
\label{eq:den_p}
\end{split}
\end{equation}

In this way, we can calculate the rms radius using
two-body densities defined in Refs. \cite{Griesshammer:2020ufp,Griesshammer:2024twu}
and the influence of SRG transformations on the 
observables can be restored by performing the unitary transformation on the two-body
density matrix. 
Since we are working with the two-body operator, the flow equation 
of the unitary transformation Eq.~\eqref{eq:srg} is also evolved at the two-body level
neglecting the induced three- and higher-body part.
It is worth noting that making the unitary transformation on this 
transitional density also allows us to study observables with non-zero momentum transfers \cite{Sun2025}.

In practice, the densities from NCSM calculations show a significant dependence on the HO basis size (determined by the maximal HO excitation $N_{\rm HO}$) and its frequency $\omega$ in both short-range ($r<1.5$~fm) and long-range parts ($r$ larger than about $3.5$~fm) regions (see Fig. 3 of Ref. \cite{Cockrell:2012vd} and Figs. \ref{fig:He4} and \ref{fig:Li6} in this work). 
The long-range observables including $\langle r^2 \rangle $ are only slightly influenced by the short-range part of the wave functions. 
The NCSM density at large distances exhibits Gaussian asymptotic behavior. Only at intermediate distances, it is characterized by physically expected $e^{-\kappa r}$ behavior. 
The size of the intermediate interval gradually enlarges with increasing basis size.
To achieve convergence for long-range operators, it is feasible to reasonably determine the long-range density tail from calculations using various basis sizes.
To address this,
we fit an exponential function $ \alpha e^{-\kappa r}$ to the density distributions obtained from the NCSM (for each $\omega$ and $N_\text{HO}$) over a specific range determined by two parameters ($r_1$ and $r_2$). The `improved' density is then defined as
\begin{equation}
\rho(r)= 
\begin{cases}
\text{NCSM density} & r \leq r_2, \\
\alpha \exp \left(-\kappa r\right) & r>r_2 .
\end{cases}
\label{eq:den}
\end{equation}
The parameters $\alpha$ and $\kappa$ 
are carefully determined by ensuring that the densities for different $N_\text{HO}$ at a given $\omega$ converge to the same values at large distances,
which will be introduced later.

The newly improved densities are then again normalized according to Eq.~(\ref{eq:dennorm}). 
This method aims to reconcile the density distributions and achieve reliable long-range behavior in NCSM calculations.

\section{Results and discussions} \label{Sec:III}
In this work, we use the chiral  
semilocal momentum-space (SMS) regularized 
NN potential at the order N$^{4}$LO$^{+}$ 
with momentum cutoffs $\Lambda_N =450\, \text{MeV}$
and the 3NF at  N$^{2}$LO 
(SMS~N$^{4}$LO$^{+}$(450)+N$^{2}$LO),
which yields a generally good description of the binding 
energies for light nuclei \cite{Maris:2020qne,LENPIC:2022cyu}.
The uncertainties on the binding energies of 
light nuclei have been discussed in Ref.~\cite{Maris:2023esu}.
The SRG flow parameter is taken to be $\lambda=1.88~\mathrm{fm}^{-1}$.
$^{4}$He is a well-bound light nucleus with a large binding energy per nucleon. 
Therefore, it is very compact 
and J-NCSM calculations converge particularly fast  for both,  energies and matter radii
leading to a mild $\omega$ and $N_\text{HO}$  
dependence. 
Therefore, it is an example nucleus to check the validity of the above-mentioned method.

\begin{figure*}[htbp]
\includegraphics[width=.8\linewidth]{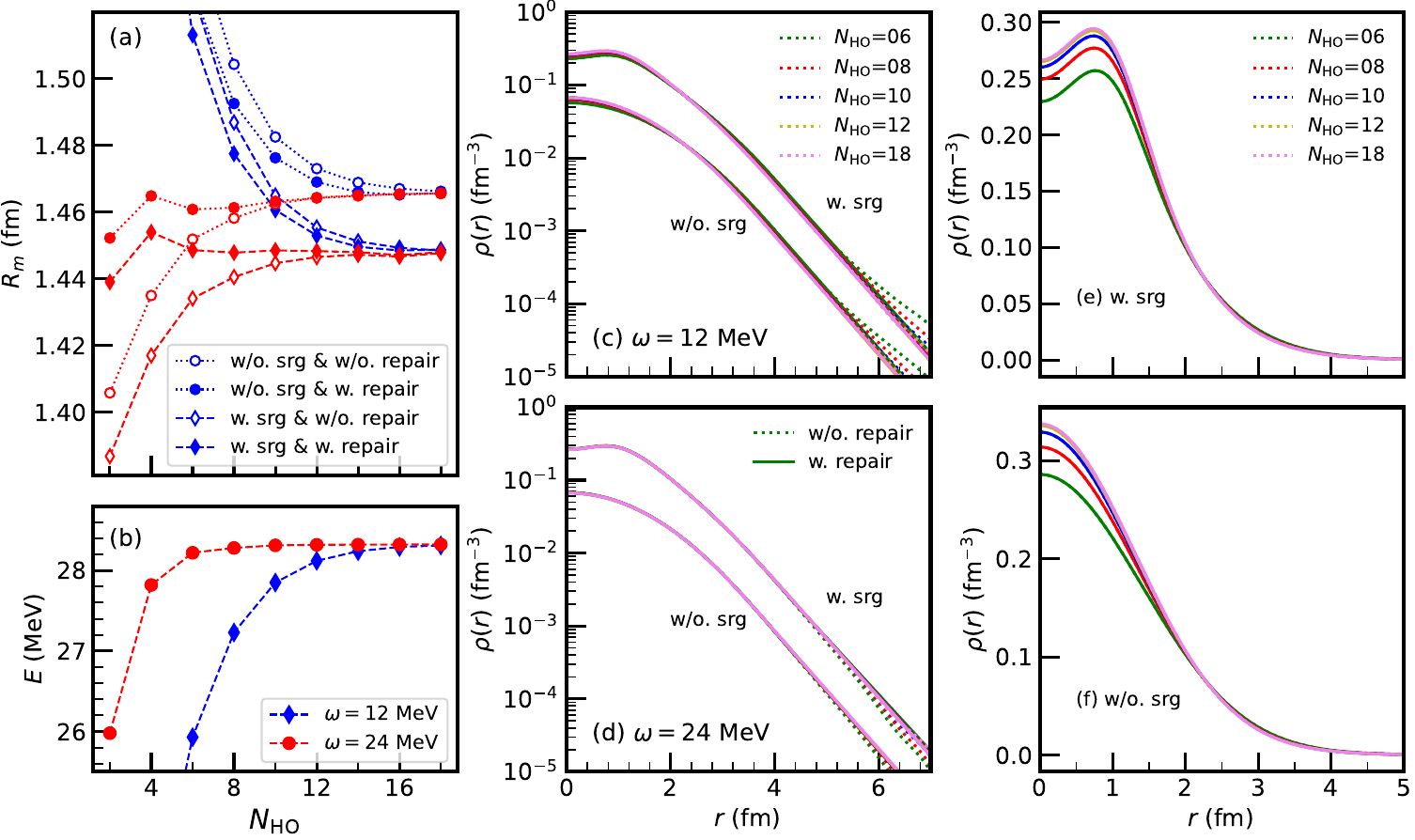}
\caption{ Ground state rms matter radii (a) and energies (b) of $^{4}$He 
as a function of the size of HO basis space for two HO frequencies 
$\omega =12$~MeV (red lines) and $24$~MeV (blue lines) calculated using the 
SMS~N$^{4}$LO$^{+}$(450)+N$^{2}$LO interaction. 
The two-body relative density distributions with or
without considering the SRG transformation in selected
basis sizes are also shown in (c) for $\omega=12$~MeV and 
(d) for $\omega=24$~MeV. 
In (c) and (d), the densities without improving the tail part are shown by
dotted lines and solid lines for those with corrections.
All densities without SRG evolution are scaled by a factor of 1/5. 
(e) and (f) for density with and without the SRG transformation for $\omega=12$ MeV but in linear scale.
}
\label{fig:He4}
\end{figure*}

In Fig. \ref{fig:He4}, 
we present our J-NCSM results for $^{4}$He.
The ground state energies illustrated 
in Fig. \ref{fig:He4}(b) exhibit 
convergence at $N_\text{HO} = 6$ for $\omega=24$~MeV 
and at $N_\text{HO} = 14$ for $\omega=12$~MeV. 
The rms matter radii indicated by the open circles 
in Fig. \ref{fig:He4}(a), 
converge at $N_\text{HO} = 12$ for $\omega=24$~MeV 
and tend to be converged at $N_\text{HO} = 18$ for $\omega=12$~MeV. 
To assess the influence of the SRG evolution on the matter radius, 
we also include results obtained by applying 
a unitary transformation to the J-NCSM wave functions 
\cite{Sun2025} obtained from the calculations with the SRG softened interaction, 
as indicated by the open diamonds in Fig. \ref{fig:He4}(a). 
The converged matter radius is found to be $1.447$~fm 
with SRG consideration and $1.466$~fm without it, 
both values being consistent with the 
experimental measurement of $1.457\pm 0.010$~fm \cite{Tanihata:2013jwa}. 
The impact of the SRG transformation on the radii is fairly small, approximately $0.02$~fm for $^{4}$He, 
being consistent with the conclusion in Refs.~\cite{Schuster:2014lga,Miyagi:2019bkl}.
This is attributed to the fact that
the expectation value of $\langle r^2 \rangle$ 
is primarily determined by the long-range behavior of the densities, 
whereas the effects of the unitary transformation
are localized to the short-range part, 
leaving the long-range part of the two-body relative density largely unaffected.
In Figs. \ref{fig:He4}(e) and (f), 
it is evident that the two-body relative densities after
considering the unitary transformation initially
increase with relative distance before damping to zero and 
the maximum is located at $r \approx 1$~fm, 
reflecting the short-range repulsive core of the two-nucleon interaction \cite{Carlson:2014vla}. 
More details on the SRG transformation on $^4$He
in our J-NCSM calculation can be found in Appendix~\ref{APP:A1}.

A comparison of densities at small basis sizes
with those at $N_\text{HO}=18$ reveals that
the long-range part does not exhibit
the correct asymptotic behavior due to the small basis size
and the limitation of the HO basis radial wave functions, 
which display Gaussian asymptotic behavior, i.e., 
decaying as $e^{-\beta r^2}$. This can be better seen 
on the logarithmic scale in Fig.~\ref{fig:He4}(c) and (d). 

By checking the densities for a fixed
$\omega$, we find that there is a crossing point, $r \approx 4$~fm, for different $N_\text{HO}$ and around this point, there is an interval in which the densities only slightly depend on the basis size, for example about $2$--$4$~fm for $\omega=24$~MeV. 
The densities in this interval for each $N_\text{HO}$ are used to determine the two parameters $\alpha$ and $\kappa$ in Eq.~(\ref{eq:den}). As described in Appendix~\ref{APP:A2}, we select from this group fits 
for which the long range densities are similar for 
different values of $N_{\rm HO}$. This procedure 
guarantees that the convergence only affects the 
short range part of the interaction which we need to 
obtain directly from the J-NCSM solutions. The trivial 
long distance behavior is fixed by our correction  
procedure. After correction, 
the densities at various HO basis sizes 
fall off as $e^{-\kappa r}$ in the long-range part, 
and 
the resulting radii are quickly converging in both 
cases, with
or without SRG transformation on the densities.
It is reassuring to see that the converged values 
agree with and without correction procedure. 
The results for $^{4}$He demonstrate that by correcting the tail part of the densities from J-NCSM calculations,
the radius converges more rapidly and confirms 
the effectiveness of density corrections in enhancing the accuracy of J-NCSM predictions for nuclear radii.

\begin{figure}[htbp]
    \centering
    \includegraphics[width=\linewidth]{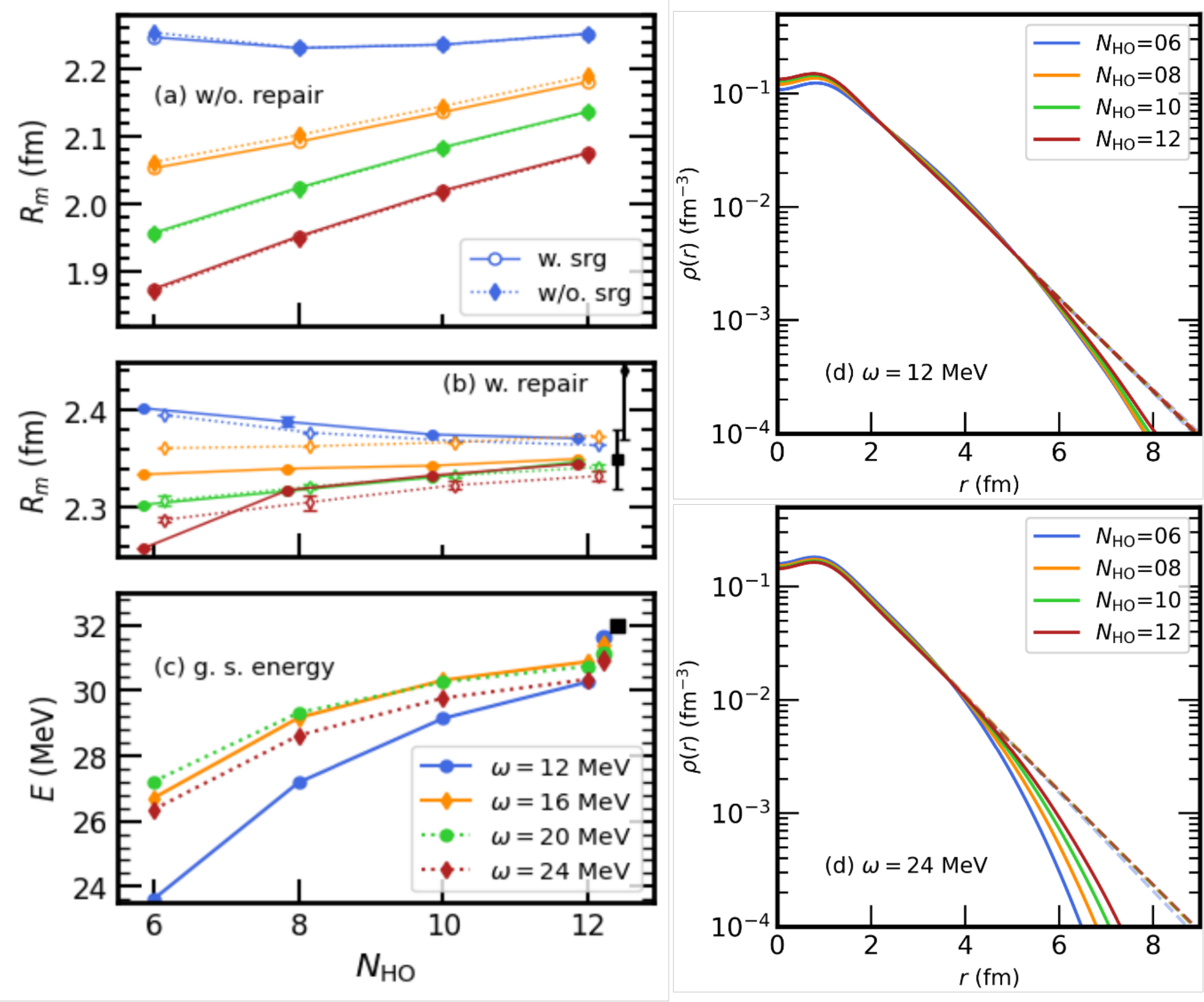}
\caption{Ground state rms matter radii (a) before correction and (b) after correction, and (c) energies of $^{6}$Li 
as a function of the size of HO basis space for four selected HO frequencies 
$\omega =12$, $16$, $20$, and $24$~MeV calculated using the 
SMS N$^{4}$LO$^{+}$ ($450$~MeV) + N$^{2}$LO interaction. 
Experimental matter radii $2.35\pm 0.03$~fm \cite{Tanihata:2013jwa} (black square)
and $2.44\pm 0.07$~fm \cite{Dobrovolsky:2006jco} (black diamond) are also shown in (b).
In (c), the black square is the experimental binding energy \cite{Wang:2021xhn} and
the additional solid points are the extrapolated values.
The two-body relative density distributions considering the SRG transformation 
(d) for $\omega=12$~MeV and 
(e) for $\omega=16$~MeV.  
The solid lines are the densities without correction and the dashed lines for these after corrections.
}
\label{fig:Li6}
\end{figure}

Then we applied the procedures as mentioned above 
to the $p$-shell helium and lithium isotopes, $^{6,8}$He and $^{6,7,8}$Li, 
whose ground-state properties including energies and radii have been
extensively studied in various NCSM calculations with
realistic forces 
\cite{Caurier:2005rb,Nogga:2005hp,
Forssen:2009vu,Cockrell:2012vd,
Bacca:2012up,Papadimitriou:2011jx,
Maris:2012bt,Romero-Redondo:2016qmc,
Forssen:2017wei,Gnech:2020qtt,Rodkin:2022fus}. 
The direct comparison of these results is not possible  
since different interactions and strategies for extracting the radius 
have been employed. It will nevertheless be instructive to 
relate our results to these benchmark data later on. 

The detailed results for an $N=Z$ nucleus $^{6}$Li are 
presented as an example of extracting the converged radii from J-NCSM calculations. 
In Figs. \ref{fig:Li6} (a) and (c), 
we display the J-NCSM calculations for the radii and energies 
for selected HO frequencies and different basis sizes from $N_\text{HO}=6$ to 12. 
It has been shown in Ref. \cite{LENPIC:2022cyu} that
the binding energies of light nuclei in question can be well
described with SMS N$^{4}$LO$^{+}$ ($450$~MeV) + N$^{2}$LO interaction. 
Here, to estimate the ground state energy,
we just employ the commonly used three-parameter formula
$E(N_\text{HO},\omega) = E(\omega)_{N_\text{HO}\rightarrow \infty}+Ae^{-BN_\text{HO}}$ and not discuss the uncertainties caused by chiral expansion and extrapolation. 
The extrapolated energies of different $\omega$ values vary from $30.8$ to $31.6$~MeV, 
being consistent with the experiment $31.9$~MeV \cite{Wang:2021xhn}. 
Regarding the radius, as observed in other NCSM calculations \cite{Shin:2016poa},
it is highly dependent on both the basis size and frequencies. 
The calculated results do not converge within the limited basis sizes, as shown in Fig. \ref{fig:Li6} (a).  
This dependence is also reflected in the long-range part of the density distributions,
which are strongly influenced by the basis size, as illustrated in Figs. \ref{fig:Li6} (d) and (e).
Especially for $\omega=12$~MeV, 
the calculated radii in Fig. \ref{fig:Li6}(a) are almost independent on the basis size but their densities are sensitive to $N_\text{HO}$ in tail parts and at the origin. 
Based on the densities from the J-NCSM, we modify the tail part,
as shown by the dashed lines in  Figs. \ref{fig:Li6} (d) and (e)
as described in Appendix \ref{APP:A2} in more detail.
The calculated radii for selected $\omega$ values 
tend to converge to the same value with increasing $N_\text{HO}$.  
As shown in Fig. \ref{fig:Li6}(b),  
the results in $N_\text{HO}=12$ for four selected  $\omega$ values range from $2.33$~fm to $2.39$~fm, 
which is consistent with the experimental value $2.35\pm 0.03$~fm \cite{Tanihata:2013jwa} 
and slightly smaller than $2.44\pm 0.07$~fm \cite{Dobrovolsky:2006jco}. 
The results from Ref.~\cite{Wolfgruber:2023ehw} based on the same 
interaction and using an ANN for extrapolation is $2.291(18)$~fm. 
Interestingly, our values are slightly larger but still marginally consistent. These results underscore the necessity of density corrections 
for achieving convergence in the radius calculations of $p$-shell nuclei within the J-NCSM framework.

\begin{figure}[tb]
    \centering
    \includegraphics[width=0.49\linewidth]{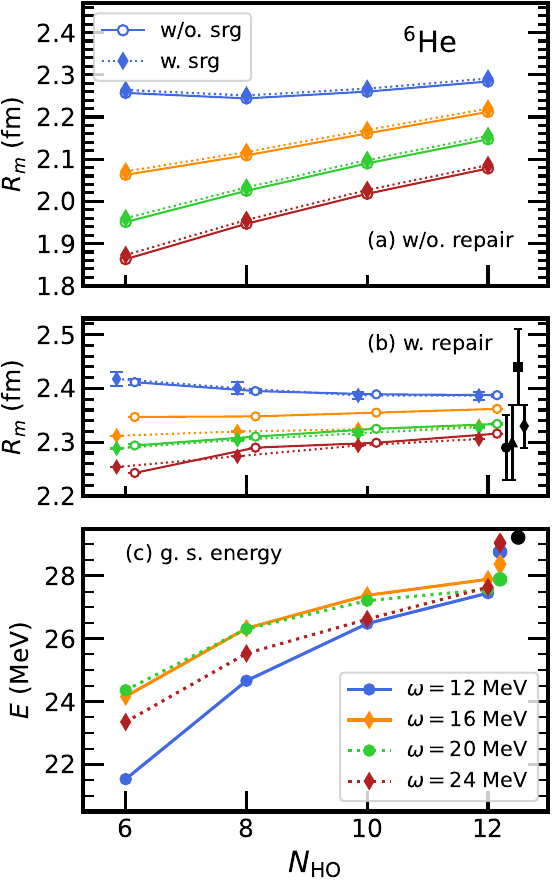}
    \includegraphics[width=0.49\linewidth]{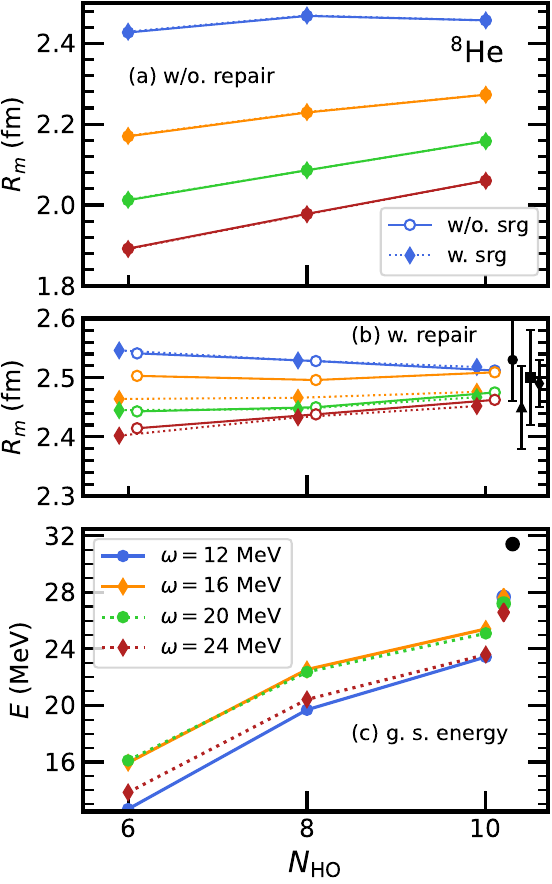}
    
    \includegraphics[width=0.49\linewidth]{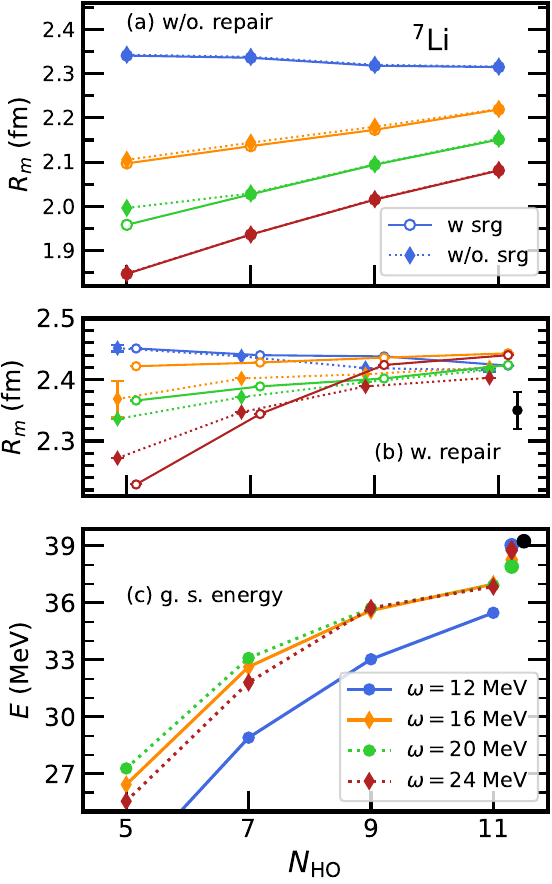}
    \includegraphics[width=0.49\linewidth]{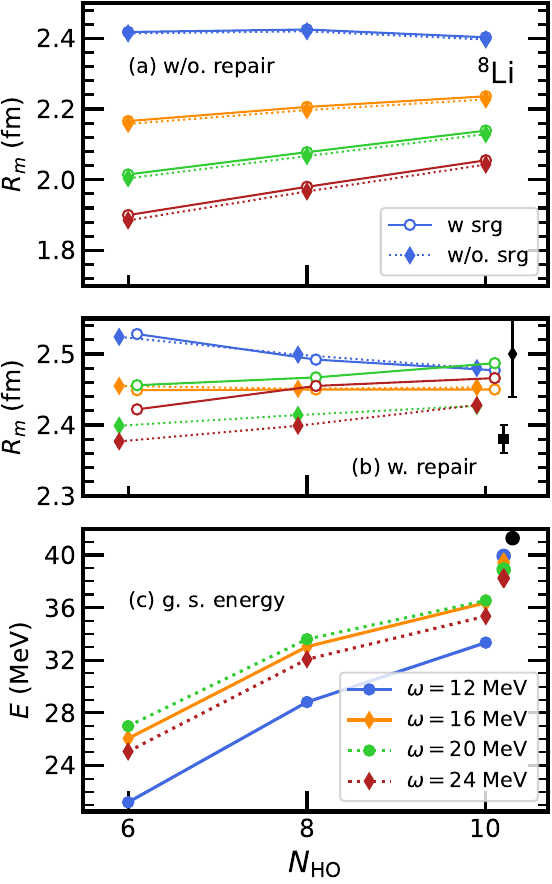}
\caption{Ground state rms matter radii before correction (a) and after correction (b), and energies (c) for $^{6,8}$He (upper left and right) and $^{7,8}$Li (lower left and right)
as a function of the size of HO basis space for four selected HO frequencies 
$\omega =12$, $16$, $20$, and $24$~MeV calculated by using the 
SMS N$^{4}$LO$^{+}$ ($450$~MeV) + N$^{2}$LO. 
The available experimental data are also shown for comparison.
}
    \label{fig:He-Li}
\end{figure}

\begin{figure}[tb]
\includegraphics[width=0.8\linewidth]{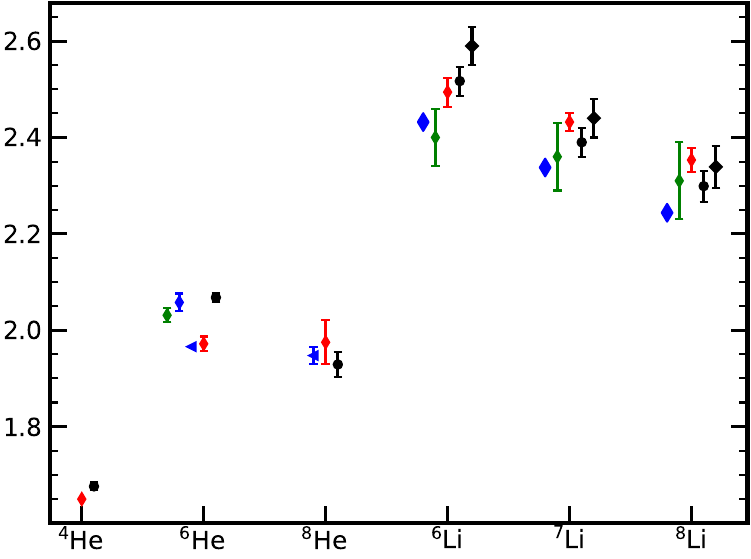}
\caption{Calculated charge radii of $^{4,6,8}$He and $^{6,7,8}$Li (red points) after improving the density tails and the comparison with available experimental data (black points taken from Ref. \cite{Tanihata:2013jwa} and black diamonds taken from Ref. \cite{Nortershauser:2011zz}).
Theoretical calculations for Li isotopes  are taken from Ref.~\cite{Cockrell:2012vd} (blue diamonds) based on the NCSM with JISP16 interaction and 
 from Ref.~\cite{Forssen:2009vu}(green diamonds)  based on the CDB2k interaction. Blue triangle points of $^{6,8}$He are proton radii from Ref.~\cite{Caprio:2014iha}.  
The blue and green solid points for $^6$He are taken from Refs.~\cite{Quaglioni:2017vpa} based on an older  SRG chiral interaction and \cite{Rodkin:2022fus} using 
the  Daejeon16 interaction.}
\label{fig:rch}
\end{figure}

The results for other $p$ shell nuclei $^{6,8}$He and $^{7,8}$Li are shown in Fig. \ref{fig:He-Li}. 
In our J-NCSM calculations,  
the extrapolated energies are mostly consistent with the data \cite{Wang:2021xhn}. 
For $^{6}$He, a two-neutron halo nucleus, our extrapolated energies ground state energies are 
$27.89$--$29.05$~MeV and the experimental binding energy is $29.22$~MeV \cite{Wang:2021xhn}. 
The calculated radii with improved densities, 
$2.32$~fm--$2.40$~fm for different $\omega$ at $N_\text{HO}=12$,
are comparable with the experimental data 
$2.29(6)$~fm \cite{Liu:2021cbn},  
$2.30(7)$~fm \cite{Alkhazov:1997zz}, 
$2.44(7)$~fm \cite{Chung:2015sza}, 
and $2.33(4)$~fm \cite{Tanihata:1992wf}.  
As mentioned earlier, a direct comparison to results using other interactions has to be taken with care. However, the results of 
Refs.~\cite{Rodkin:2022fus} and \cite{Caprio:2014iha} using 
the JISP16 interaction of  $2.342(7)$~fm and $2.32$~fm  
are consistent with our values. However, our value is slightly 
smaller than theoretical calculations with the NCSM in continuum $2.46(2)$~fm \cite{Quaglioni:2017vpa} using the Idaho~N$^3$LO interaction \cite{Entem:2003ft} 
 SRG-evolved without adding a 3NF. 
For $^{8}$He, the extrapolated energies are smaller than the experimental 
results. This was already observed in Ref.~\cite{LENPIC:2022cyu}. 
In fact, using a lower order NN interaction \cite{Maris:2020qne}, one obtains better agreement with experiment. 
Therefore, it will be interesting to see  
predictions for this nucleus using additional higher-order 3NFs. 
However, our calculated radii with the corrected densities are  consistent 
with the data
$2.53(7)$~fm \cite{Liu:2021cbn}, 
$2.45(7)$~fm \cite{Alkhazov:1997zz}, 
$2.50(8)$~fm \cite{Chung:2015sza},
$2.49(4)$~fm \cite{Tanihata:1992wf},
all of which have large uncertainties because the matter radius is usually deduced from cross-section data, 
relying on the adopted asymptotic behavior of the density tail.
The calculations for ground state energies of $^{7,8}$Li are in agreement with the experimental data.
For the matter radii of Li isotopes, there are two sets of experimental data in Ref.~\cite{Dobrovolsky:2006jco}  and Ref.~\cite{Tanihata:1992wf}. 
The former \cite{Dobrovolsky:2006jco} shows relatively large matter radii and our results, after improving the tail parts, are close to them. 
Generally speaking,
the results with corrections on densities for different $\omega$ values can reach almost the same values with increasing basis sizes. 
The comparison to the ANN extrapolated 
values of Ref.~\cite{Wolfgruber:2023ehw} using the same interaction 
is quite interesting. 
Their results for the rms matter radius of $^7$Li and $^8$Li are $2.325(14)$ and $2.327(14)$~fm which is visibly smaller than our results of 
$2.401$-2.443~fm and $2.424$-2.480~fm. Similar as for 
$^6$Li, the correction of the long-range tail of the density also leads to an increase of the radius for the neutron-rich isotopes. 
This indicates that 
the repaired densities can help to improve predictions for radii
for the $p$-shell.

Finally, we study the charge radii. For this observable, the experimental 
uncertainties are much smaller. Therefore, other aspects like NN currents \cite{Filin:2020tcs} become more important when comparing 
to the experiment. We obtain the charge radius $R_c$ by the relation \cite{Friar1975_ANP8-219}
\begin{equation}
R_c^2 = R_p^2 + r_p^2 + \frac{N}{Z}r_n^2 + \frac{3}{4m_p^2 },
\end{equation}
in which $r_p = 0.8409$~fm \cite{ParticleDataGroup:2020ssz} (see also~\cite{Lin:2021xrc}), 
$r_n^2=0.1155$~fm$^2$ \cite{ParticleDataGroup:2020ssz}, and 
${3}/(4m_p^2) = 0.033$~fm$^2$ neglecting the NN current contributions.
It should be mentioned that the spin-orbit term \cite{Ong:2010gf}  also influences the charge radii but, due to its small contributions
\cite{Ong:2010gf}, we do not consider it in this work. 
The point-proton rms radii are calculated by 
correcting the tail behavior of the integrand
in Eq. \eqref{eq:radius}.
In Fig. \ref{fig:rch}, we show our results (red points) along with the comparison with the available experimental data and other theoretical calculations. 
The error in our result in this figure is the difference from the maximum and minimum for four $\omega$ values with the largest basis size
for each nucleus.
The error bars of our results represent the differences between the maximum and mean values for the
charge radii obtained from four selected $\omega$ with the maximal $N_\text{HO}$ for each nucleus.
For $^{6}$He, both our calculations and the results from \cite{Caprio:2014iha,Rodkin:2022fus} underestimate the experimental data, and 
the result from the NCSM in continuum \cite{Quaglioni:2017vpa} agree with the data \cite{Papadimitriou:2011jx}.
Our results for $^{8}$He has a large error bar for two reasons:
the calculated binding energy differs by about 4~MeV compared with data and after corrections for $N_\mathrm{HO}=6,8,10$ the charge radius values  still depend on $\omega$, especially a value of 2.10~fm for $\omega=12$ MeV increases the uncertainty.
The  charge radii for the other frequencies $\omega=16$, $20$ and $24$~MeV
are smaller, between $1.96$ and $2.02$~fm, and closely align  with data and with the values from Ref. \cite{Caprio:2014iha}. 
For $^{6,7,8}$Li, our results are slightly larger than other theoretical results \cite{Cockrell:2012vd,Forssen:2009vu} and are more consistent with the data.  A more careful comparison using the same interactions 
and/or NN currents is beyond the scope of this work. 
We nevertheless take the results as an indication that 
our method effectively determines the point proton radius as well
leading to a slight increase in values compared to previous work.

\section{Summary}\label{Sec:IV}
In our study, we conduct J-NCSM calculations to investigate the nuclear properties of light nuclei $^{4,6,8}$He, and $^{6,7,8}$Li
using the modern high-precision chiral two-nucleon and three-nucleon forces SMS N$^4$LO$^+$+N$^2$LO with momentum cutoff 450 MeV, focusing on ground state energies and non-converged results of matter radii. 
The rms radii are calculated using the two-body relative densities
with or without considering the influence of the SRG evolution and 
our results demonstrate that the size of a nucleus is almost not affected. 
The core idea of this work is that one can deduce the correct long-range asymptotics for the densities from those of NCSM calculations with limited basis sizes. 
Applying this idea to the matter and charge radii of $^{4,6,8}$He and $^{6,7,8}$Li proves that
correcting density tails improves the convergence of matter radii,
thus providing an alternative way to extract the radii in NCSM calculations with limited basis size.
The results underscore the importance of addressing density asymptotics and basis size effects in achieving accurate theoretical predictions that align closely with experimental observations.
One can also extend the same method to other long-range observables to check the validity of our methods. For even better accuracies, it could be interesting to use the corrected densities and ANNs along the lines of Ref.~\cite{Wolfgruber:2023ehw} to get converged results including uncertainty estimates. At the same, the transition densities that are the basis of the 
present study can be used to include NN current contributions in observables. 
Moreover,
our approach can be implemented into other \textit{ab initio} calculations based on HO 
basis functions, thus providing a potential way to address the long-standing issue of calculations with high-precision realistic nuclear forces producing accurate energy values but underestimating the size of finite nuclei.
 
\acknowledgments
X.-X. Sun thanks Shuang Zhang, Zhengxue Ren, and Harald W. Grie{\ss}hammer for helpful discussions.
This work was supported in part by the European
Research Council (ERC) under the European Union's Horizon 2020 research
and innovation programme (grant agreement No. 101018170)
as well as by the Deutsche Forschungsgemeinschaft (DFG,German Research
Foundation) and the NSFC through the funds provided to the Sino-German 
Collaborative Research Center TRR110 “Symmetries and the
Emergence of Structure in QCD” (DFG Project ID 196253076--TRR
110, NSFC Grant No. 12070131001).  
 The numerical calculations were performed on JURECA
of the J\"ulich Supercomputing Centre, J\"ulich, Germany.

\appendix
\setcounter{figure}{0}
\renewcommand\thefigure{A\arabic{figure}}
\setcounter{table}{0}
\renewcommand\thetable{A\arabic{table}}

\section{Dependence of \texorpdfstring{$^{4}$He}{4He} matter radii on the SRG flow parameter}\label{APP:A1}

We have implemented SRG transformations on the two-body relative wave functions.
Therefore, the transformation of the densities has been done on the two-nucleon 
level and three-nucleon contribitions to the SRG transformation are neglected.  
In the following, we aim at estimating the resulting uncertainty due to this 
approximation on the radius. For this, $^4$He is an especially good test case 
due to its large binding energy per nucleon. This increases 
the effects of the SRG transformation on the radius compared 
to the other nuclei in this work.

\begin{table}[tbp]
     \caption{The rms matter radii of $^4$He using the two-body relative densities with and without the SRG transformation on the J-NCSM wave function. The adopted chiral interaction is SMS N$^4$LO$^+$+N$^2$LO with momentum cutoff $400$~MeV with three flow parameters  $1.88$, $3.00$, and $4.00$~fm$^{-1}$. 
     The calculations from Faddeev-Yakubovsky (FY) with the SRG unevolved interactions are shown for comparison.}
    \begin{tabular}{ccc} 
    \toprule
           &  without &   with \\
    \midrule
         $1.88$~fm$^{-1}$&  $1.463$~fm&  $1.444$~fm\\
         $3.00$~fm$^{-1}$&  $1.437$~fm&  $1.428$~fm\\ 
         $4.00$~fm$^{-1}$&  $1.435$~fm&  $1.426$~fm \\ 
         bare(FY)      & \multicolumn{2}{c}{$1.431$~fm}\\
    \bottomrule
    \end{tabular}
    \label{tab:tab1}
\end{table}

In Table \ref{tab:tab1},
we show the rms matter radii of $^{4}$He with and without considering 
the SRG transformation. We used the  
SMS N$^4$LO$^+$+N$^2$LO interaction with a momentum cutoff of $400$~MeV and 
performed the J-NCSM calculation with $\omega=16$~MeV and $N_\text{HO}=32$.
The is close to the optimal $\omega$ and the model space is large 
enough for convergence. 
The difference in matter radii between the two cases 
decrease with increasing SRG flow parameter both converging towards the 
bare result obtained using a Faddeev-Yakubovsky (FY) calculation in momentum 
space. This is consistent with the results in Refs.~\cite{Schuster:2014lga,Miyagi:2019bkl}.
At the flow parameter of $1.88$~fm$^{-1}$ mostly used in our work, the deviation 
without SRG transformation to the bare result for the radius is $0.032$~fm$^{-1}$.  Including the 
transformation this deviation shrinks to $0.013$~fm$^{-1}$. The latter number indicates 
the missing three-nucleon contribution to this value. 

\section{Determination of radius of \texorpdfstring{$^6$Li}{6Li} }
\label{APP:A2}

$^{6}$Li with a small $np$ separation energy is a good example to test our
new procedure for determining the radius. The key point is to find 
a suitable long-range behavior determined by two parameters $\alpha$ and $\kappa$ based on the densities from limited basis sizes. 
Let us take the densities considering the SRG transformation for $\omega=16$~MeV with 
$N_\mathrm{HO}=6,8,10,12$ as examples to show how we find the improved  densities.
From the density profiles in Fig.~\ref{fig:Li6}, we get that there is a crossing point at about $r=4.2$~fm for the different $N_\mathrm{HO}$
which we take as approximation of our choice for $r_2$.
$r_1$ is determined by the crossover points between the SRG evolved 
and SRG non-evolved densities.
In the fitting process,
$r_2$ is set to be varying in [$4.2$~fm, $4.8$~fm] and [$2.5$~fm, $3.5$~fm] for $r_1$ in steps of $0.1$~fm. 
Therefore, for each $N_\mathrm{HO}$, we have $77$ different 
intervals that lead to $77$ different values of $\alpha$ and $\kappa$.
To ensure the same long-range behavior for different 
values of $N_\mathrm{HO}$,  
we select from these $4\times 77$ cases 
groups of 4 with different  $N_\mathrm{HO}$, 
for which the  densities at $r=7.0$~fm are the same 
within a given tolerance. Of course, the number of groups 
for which this condition holds decreases with decreasing 
tolerance. We have then lowered the tolerance such that the 
predicted radii for a given  $N_\mathrm{HO}$ 
are the same for each accepted case. For the example, 
this point was reach for a tolerance of $3 \cdot 10^{-6}$. 

\begin{figure*}
    \centering
    \includegraphics[width=\linewidth]{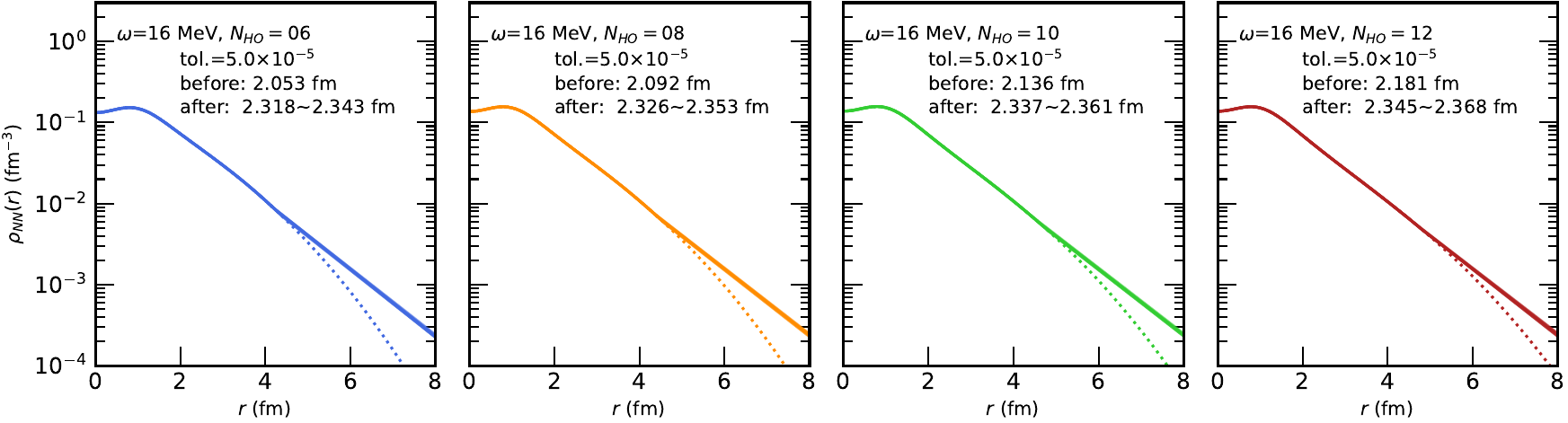}
    \includegraphics[width=\linewidth]{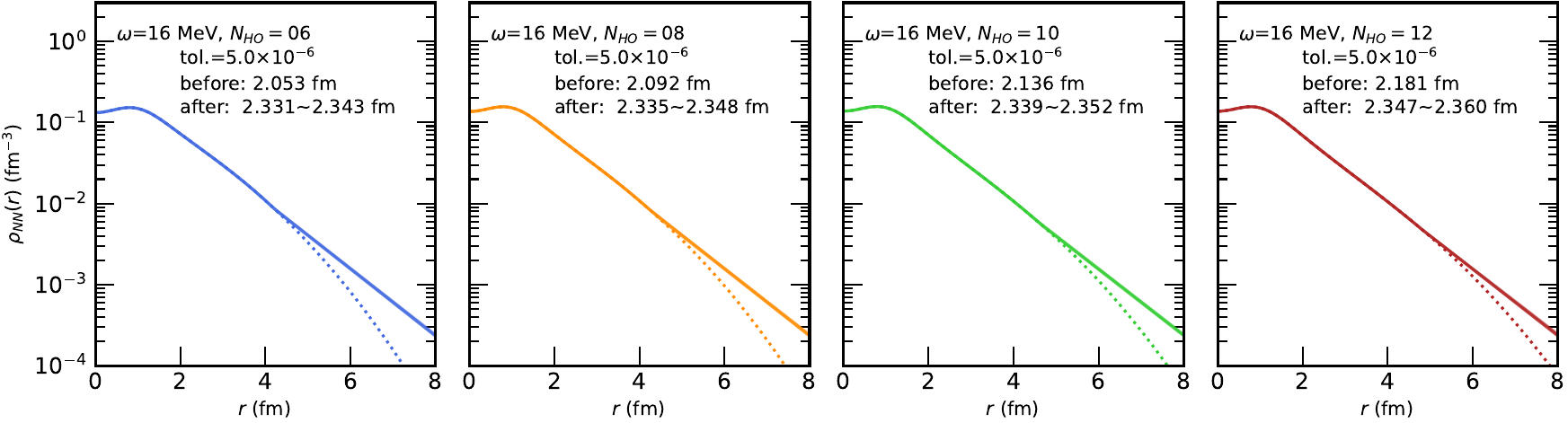}
    \includegraphics[width=\linewidth]{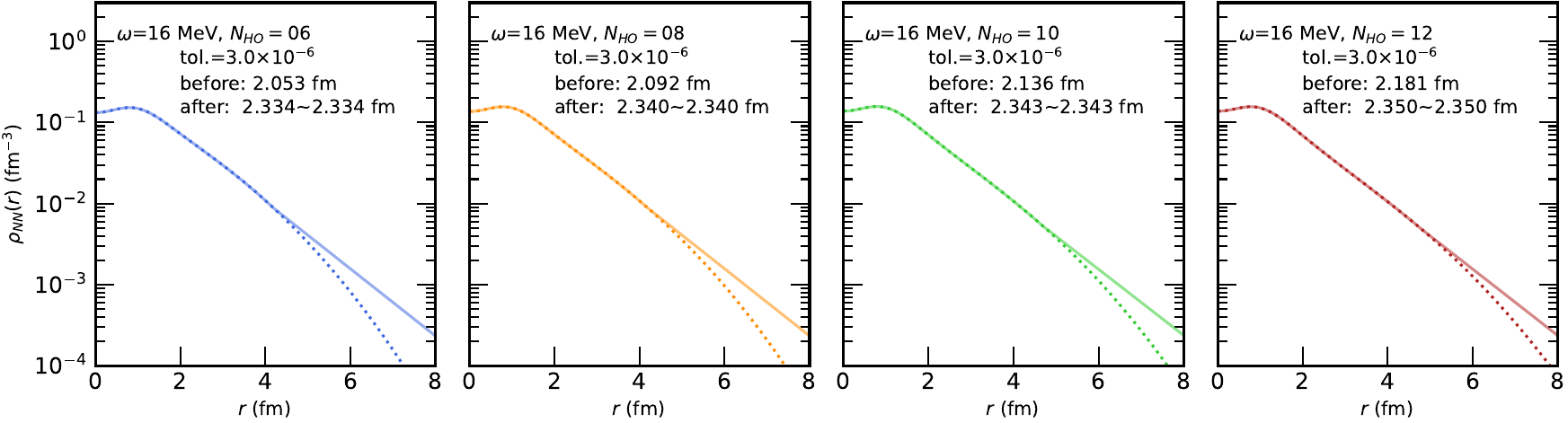}
\caption{Two-body relative densities profiles for $^6$Li for $\omega=16$~MeV.
The top panel shows densities with J-NCSM calculations and
the improved version with the tolerance of $5 \cdot 10^{-5}$ for different basis size $N_\text{HO}=6,8,10$, and 12 from left to right, the middle panel for the tolerance of $5\cdot10^{-6}$, and the bottom panel for $3 \cdot 10^{-6}$. In each subfigure, the calculated rms matter radii before and after correction are also given.
For the lowest panel, the tolerance is small enough 
to shrink the range of radii to one results for each $N_\text{HO}$. 
}
    \label{fig:Li6-A1}
\end{figure*}

In Fig.~\ref{fig:Li6-A1}, we show the densities from J-NCSM calculations (dashed lines) and the improved ones (solid lines) for $N_\mathrm{HO}=6,8,10$, and $12$ for 
all accepted cases using three difference tolerances.
As can be seen, the different corrected densities are very 
similar to each other. The range of the extracted radii goes towards a single value by construction when decreasing
the tolerance. Note that this does not imply that the radius is independent of $N_\mathrm{HO}$. The small 
variations that remain here are the result of the convergence at short distances. However, the variation 
is significantly smaller than without applying the correction and the best value is shifted from 
$2.181$~fm to $2.350$~fm which is a signifcant 
improvement. 

\begin{figure*}
    \centering
    \includegraphics[width=\linewidth]{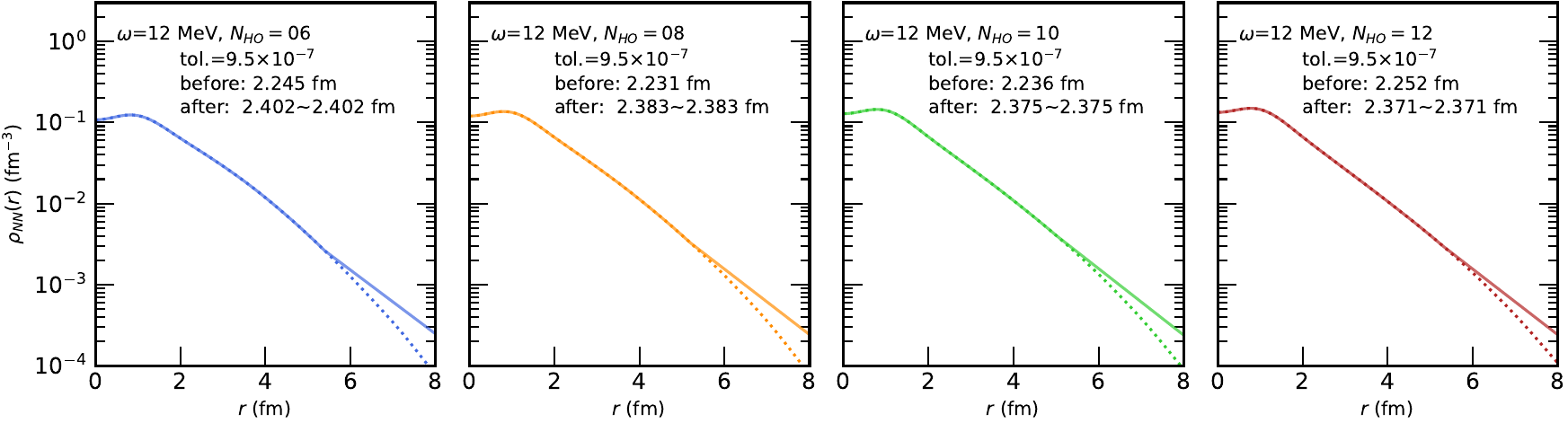}
    \includegraphics[width=\linewidth]{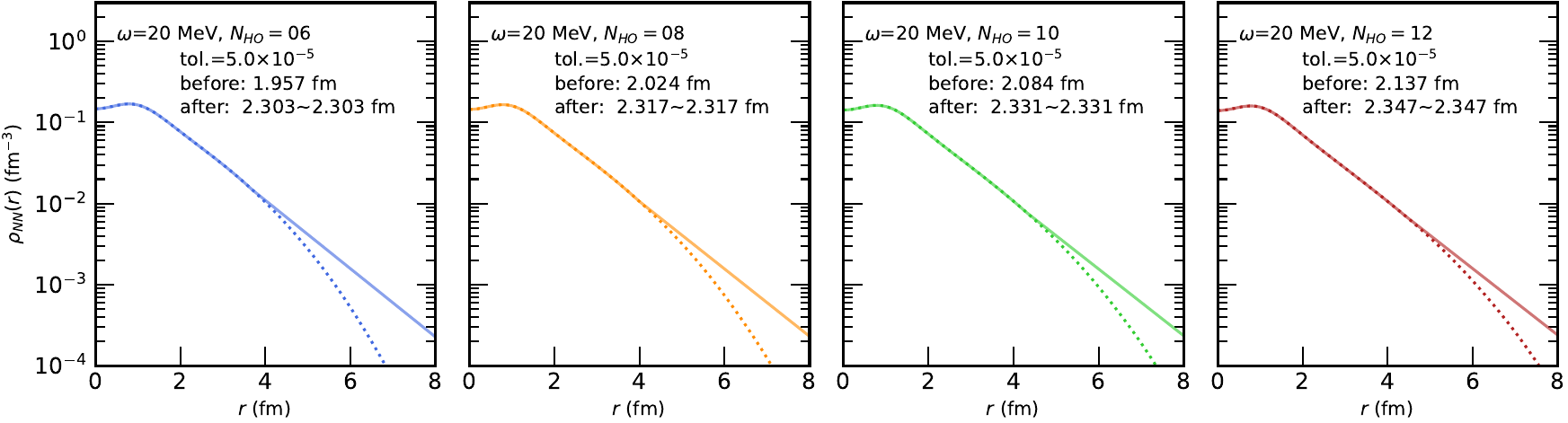}
    \includegraphics[width=\linewidth]{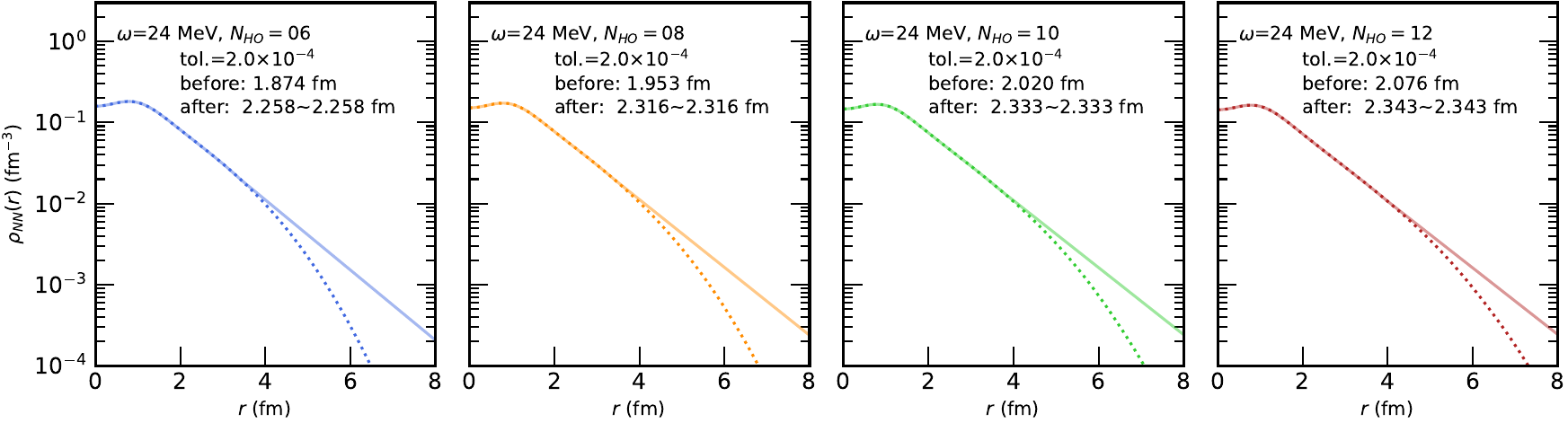}
\caption{Densities profiles for $^6$Li. The best fit for the densities from J-NCSM calculations with $\omega=12$~MeV (top panel), $20$~MeV (middle panel), and $24$~MeV (bottom panel).
In each subfigure, the adapted tolerance and the calculated rms matter radii before and after correction are also given.
}
    \label{fig:Li6-A2}
\end{figure*}

In order to assess the uncertainty of the procedure, 
we show the result of the correction procedure from 
above for different values of $\omega$ in 
Fig.~\ref{fig:Li6-A2}.
It can be seen  that after correcting the tail parts,
the dependence of the calculated matter radius on
both $N_\text{HO}$ and $\omega$ become weaker compared 
with those directly obtained from J-NCSM. The procedure 
improves the convergence pattern and results in an increased 
accuracy for the radii.

\bibliographystyle{apsrev4-2}
\bibliography{ref}

\end{document}